\documentclass[]{gPCH2e}
\usepackage{graphicx}
\usepackage{amsmath}
\usepackage{amsfonts}
\usepackage{amssymb}

\begin{document}

\jvol{00} \jnum{00} \jyear{2013} \jmonth{October}

\received{Received 31 October 2013}

\articletype{RESEARCH ARTICLE}


\title{Fingerprints of antiaromaticity in the negative ion (Li$_3$Al$_4$)$^-$
via an \emph{ab initio} quantum-chemical study of the equilibrium structure of
the inhomogeneous electron liquid}

\author{%
A. Grassi$^{a}$, 
G. M. Lombardo$^{a}$, 
G. G. N. Angilella$^{b,c,d,e\ast}$\thanks{$^\ast$Corresponding author.
Email: giuseppe.angilella@ct.infn.it\vspace{6pt}},
N. H. March$^{f,g}$,
R. Pucci$^{b,d}$,
D.~J.~Klein$^{h}$,
A. T. Balaban$^{h}$
\\\vspace{6pt} 
$^{a}${\em{Dipartimento di Scienze del Farmaco,
Universit\`a di Catania,
Viale A. Doria, 6, I-95126 Catania, Italy}};
$^{b}${\em{Dipartimento di Fisica e Astronomia, Universit\`a di Catania,
64, Via S. Sofia, I-95123 Catania, Italy}};
$^{c}${\em{Scuola Superiore di Catania, Universit\`a di Catania, Via
Valdisavoia, 9, I-95125 Catania, Italy}};
$^{d}${\em{CNISM, UdR Catania, 64, Via S. Sofia, I-95123 Catania, Italy}};
$^{e}${\em{INFN, Sez. Catania, 64, Via S. Sofia, I-95123 Catania, Italy}};
$^{f}${\em{Department of Physics, University of Antwerp,
Groenenborgerlaan 171, B-2020 Antwerp, Belgium}};
$^{g}${\em{Oxford University, Oxford, UK}};
$^{h}${\em{MARS Department, Texas A\&M University at Galveston, Galveston, TX 77553-16756, USA}}.}

\maketitle

\begin{abstract}
Fingerprints of antiaromaticity in the negative ion (Li$_3$Al$_4$)$^-$, this
species being realizable via a laser vaporization technique, are revealed by
means of an \emph{ab initio} quantum-chemical investigation. First, the
ground-state equilibrium geometry of this ion is predicted. Also, for the
corresponding inhomogeneous electron liquid, the
characteristics of the HOMO are studied, both for the square and the rectangular
Al$_4$ geometry in two low-lying isomers of the negative ion. There is no
particular sensitivity to the change in geometry of the Al$_4$ configuration.
Therefore, we have calculated theoretically chemical shifts (NICS), which contain
remarkable fingerprints of antiaromaticity. As to future directions, some
comments are added in relation to the Shannon entropy.

\medskip
\end{abstract}

\section{Background and outline}

The present study uses \emph{ab initio} quantum-chemical computations to treat
(Li$_3$Al$_4$)$^-$, and then looks into possible ``aromatic'' characteristics,
as has become a popular idea for metallic species \cite{Boldyrev:05,Chen:05}.
The general idea of aromaticity goes back well over a century, and was centrally
important in Pauling and Wheland's explication \cite{Pauling:33a} of the special
stability of benzene, where the particular characteristic focused on was the
thermodynamic stabilization (or resonance) energy.  Over the years a variety of
experimental manifestations of aromaticity have been discussed: exceptional
kinetic stability; exceptional thermodynamic stability; exceptional geometric
structure (bond lengths intermediate between single and double, \emph{e.g.,} for
benzene); exceptional polarizabilities; exceptional magnetic susceptibilities
(and ring currents); and exceptional chemical shifts.  See, \emph{e.g.,} a
serious monograph \cite{Minkin:94}, or a special issue of Chemical Reviews
\cite{Schleyer:01}, and also Randic's review \cite{Randic:03}.  Of these
different manifestations, that of anomalous chemical shifts, has recently become
of great interest, when Schleyer \cite{Schleyer:96} realized that theoretically
one could readily compute the chemical shift even without a real nucleus placed
in the vicinity of the aromatic region, whence nucleus-independent chemical
shifts (NICS) have become a standard means to test for aromaticity (see also
Kuznetsov \emph{et al.} \cite{Kuznetsov:03}).  But generally ``aromaticity'' is
properly a correlated
\cite{Katritzky:98,Klein:92,Klein:97,Katritzky:01,Cyranski:02,Balaban:04} manifestation the
whole collection of the above noted features, though then the degree of
correlation between these different manifestations becomes of question ---a
question which is not theoretically fully resolved, especially for aromaticity
outside of the classical case of conjugated carbon networks.  Thus, further
testing of the aromaticity idea in this more general context then is of general
interest, particularly as the understanding of novel metal clusters is so
interesting itself \cite{Clayborne:10}.

Of particular relevance for the present Letter is Section~7 of Boldyrev and
Wang's review \cite{Boldyrev:05} referring to (Li$_3$Al$_4$)$^-$, as well as to
the Schleyer group's work \cite{Chen:05} on the Al$_4^{2-}$ dianion.  We utilize
Schleyer \emph{et al.}'s \cite{Schleyer:96} nucleus-independent chemical shift
criterion (NICS) for aromaticity.

With this as background, the outline of our present work is as follows. In
Section~\ref{sec:qcp} immediately below, we record quantum-chemical predictions
of the geometries of two low-lying isomers of (Li$_3$Al$_4$)$^-$. We also carry
out there a stability test of these two geometries, which is positive, the
normal mode vibrational frequencies being tabulated, as well as other HOMO
properties of these isomers. Section~\ref{sec:shifts}, which with the geometries
in Fig.~\ref{fig:1} is at the heart of the present study, deals quantitatively
with chemical shifts, which are shown to contain very pronounced fingerprints of
antiaromaticity in (Li$_3$Al$_4$)$^-$. 

\section{Quantum-chemical prediction of geometries of low-lying isomers of
(Li$_3$Al$_4$)$^-$}
\label{sec:qcp}

\begin{table}[t]
\begin{center}
\begin{tabular}{ll}
\multicolumn{1}{c}{Anti-aromatic} & \multicolumn{1}{c}{Aromatic} \\
\multicolumn{1}{c}{(Fig.~\ref{fig:1}a)} &
\multicolumn{1}{c}{(Fig.~\ref{fig:1}b)} \\
125.0279 & 123.0215 \\
140.9736 & 152.2599 \\
153.7713 & 167.7396 \\
157.0725 & 178.0138 \\
188.7271 & 208.0585 \\
220.1328 & 217.0149 \\
233.7731 & 224.6717 \\
234.4931 & 260.3682 \\
239.0553 & 265.0598 \\
299.0800 & 266.4721 \\
323.5899 & 296.6529 \\
330.2183 & 336.9039 \\
341.5039 & 353.8490 \\
361.6591 & 357.7507 \\
375.4484 & 373.3925
\end{tabular}
\end{center}
\caption{Normal mode vibrational frequencies (in a.u.) for both isomers shown in
Fig.~\ref{fig:1}.}
\label{tab:1}
\end{table}

The quantum-chemical ab-initio calculations have been performed using the
Gaussian 09W package \cite{Frisch:09a}. All the structures were energy optimized
with second order M\o{}ller-Plesset (MP2)
\cite{Moeller:34,HeadGordon:88,Saebo:89,Frisch:90,Frisch:90a,HeadGordon:94}
perturbation method, considering all the valence and core electrons (MP2=FULL)
at the 6-311G \cite{McLean:80,Raghavachari:80} level of the theory, adding
polarization (d,p) \cite{Frisch:84} and diffuse (++) \cite{Clark:83} functions.
Local minima were attained with the program's default geometry optimization
(Berny algorithm) \cite{Xi:06} procedure and settings.  To obtain a square
planar structure for the four Al atoms, we constrained the four Al-Al distances
to be equal with the valence angles at $90.0^\circ$. Nuclear magnetic resonance
(NMR) Shielding Tensors were obtained with the Gauge-Independent Atomic Orbital 
(GIAO) method \cite{Cheeseman:96} for the optimized MP2 structures.

Fig.~\ref{fig:1} shows the geometry thereby predicted for two low-lying isomers
of the negative ion under discussion. Fig.~\ref{fig:1}a we have anticipated to
have the anti-aromatic species, with the Al nuclei on the vertices of a slightly
distorted rectangle. In contrast, Fig.~\ref{fig:1}b shows the Al$_4$ square
geometry, the common bond-length being $\sim 2.637$~\AA.

As a stability test, we record in Table~\ref{tab:1} the normal mode vibrational
frequencies for both isomers shown in Fig.~\ref{fig:1}. The overall range of
frequencies is practically the same in the two cases, though differences of
detail are in evidence. Both isomers predicted in Figs.~\ref{fig:1}a and
\ref{fig:1}b pass the stability test, all the frequencies in Table~\ref{tab:1}
being real.

\begin{figure}[t]
\begin{center}
\begin{tabular}{crcrc}
Anti-aromatic & \multicolumn{3}{c}{Eigenvalues} & Aromatic \\
\begin{minipage}[t]{0.25\columnwidth}
\begin{center}
\includegraphics[width=\textwidth]{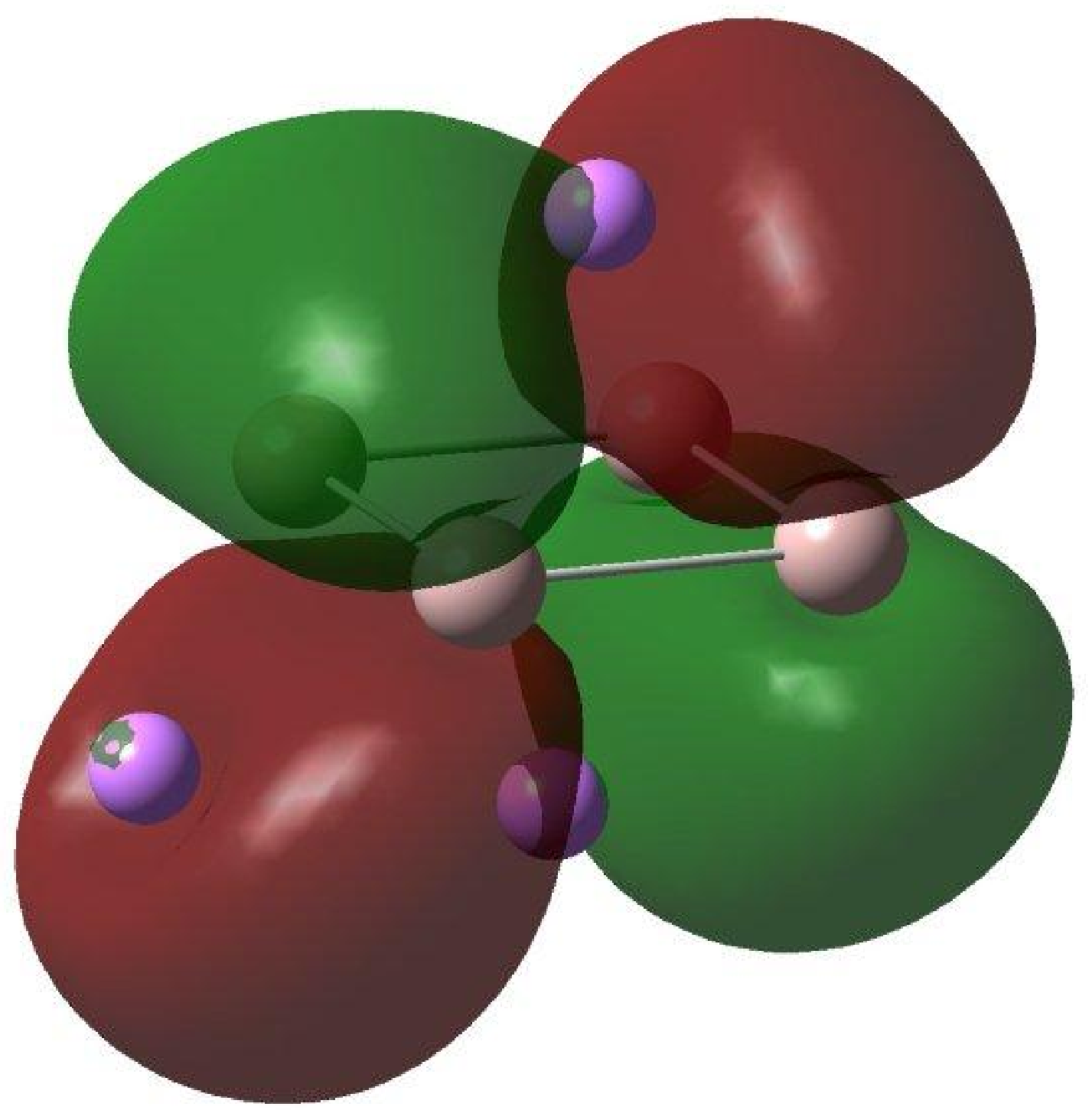}
\end{center}
\end{minipage}
& $-0.04306$ & HOMO & $-0.03683$ &
\begin{minipage}[t]{0.25\columnwidth}
\begin{center}
\includegraphics[width=\textwidth]{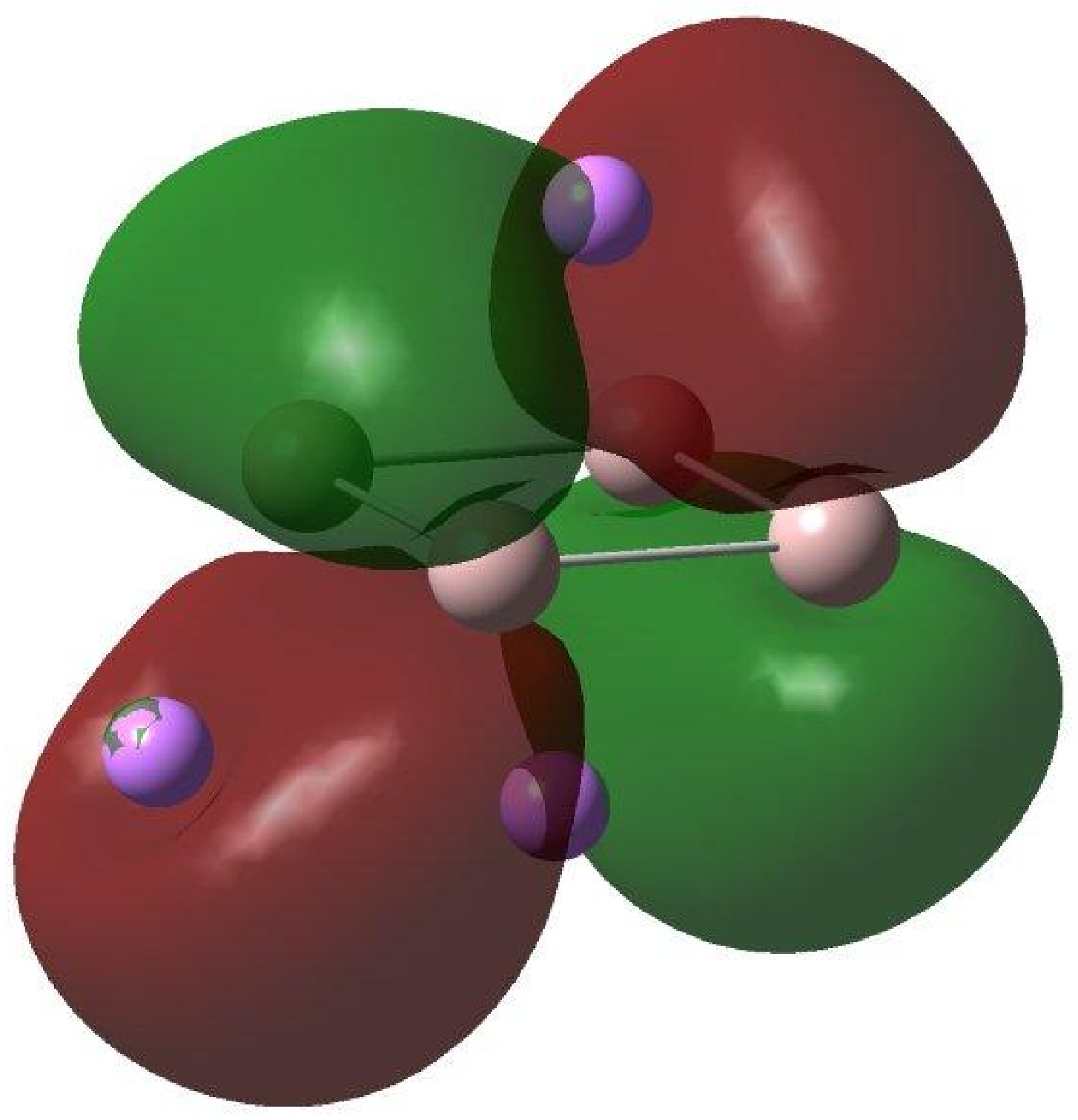}
\end{center}
\end{minipage} \\
(a) & & & & (b) \\
\begin{minipage}[t]{0.25\columnwidth}
\begin{center}
\includegraphics[width=\textwidth]{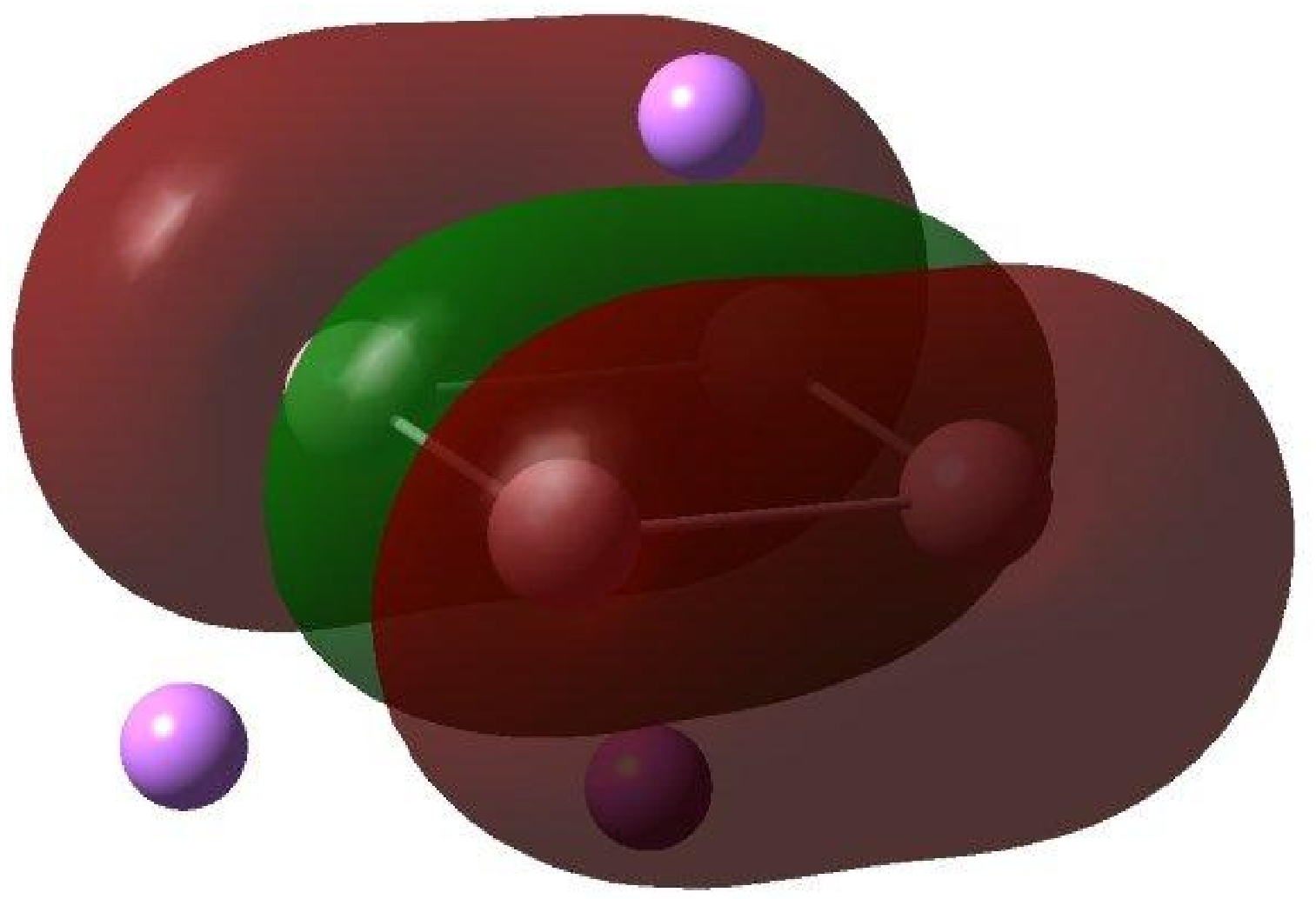}
\end{center}
\end{minipage}
& $-0.06416$ & HOMO $-1$ & $-0.07226$ &
\begin{minipage}[t]{0.25\columnwidth}
\begin{center}
\includegraphics[width=\textwidth]{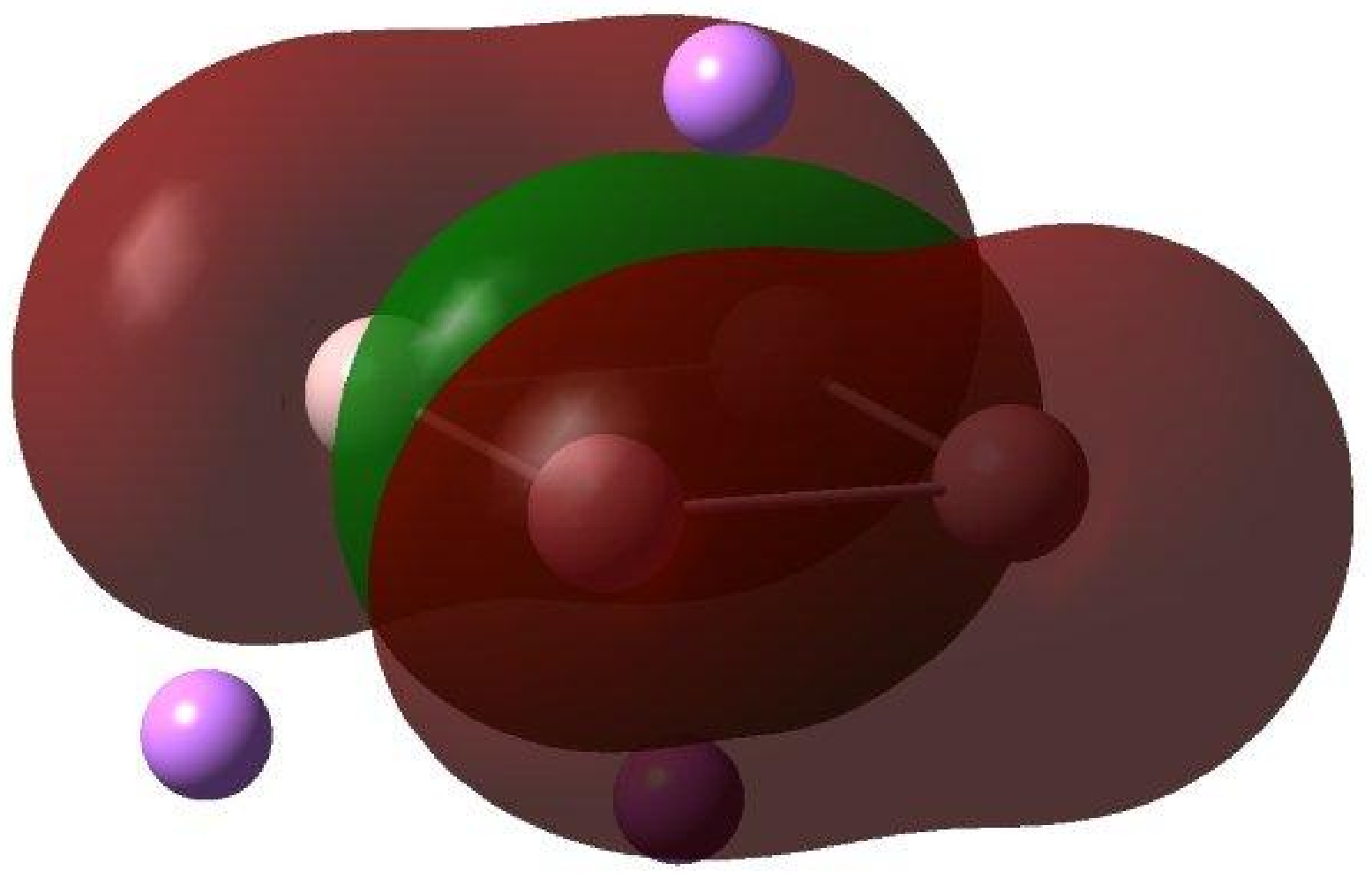}
\end{center}
\end{minipage} \\
(c) & & & & (d) \\
\begin{minipage}[t]{0.25\columnwidth}
\begin{center}
\includegraphics[width=\textwidth]{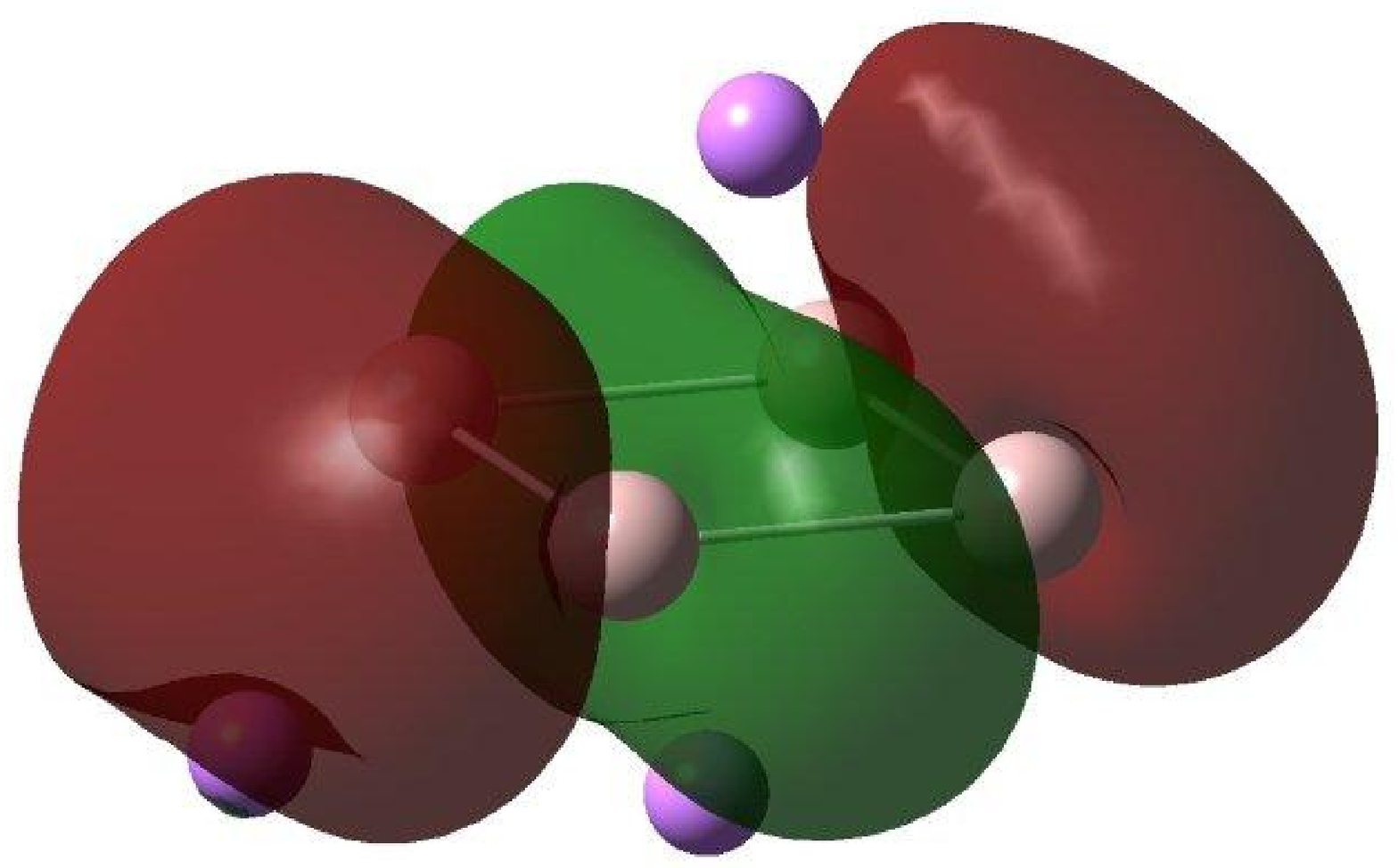}
\end{center}
\end{minipage}
& $-0.08752$ & HOMO $-2$ & $-0.08473$ &
\begin{minipage}[t]{0.25\columnwidth}
\begin{center}
\includegraphics[width=\textwidth]{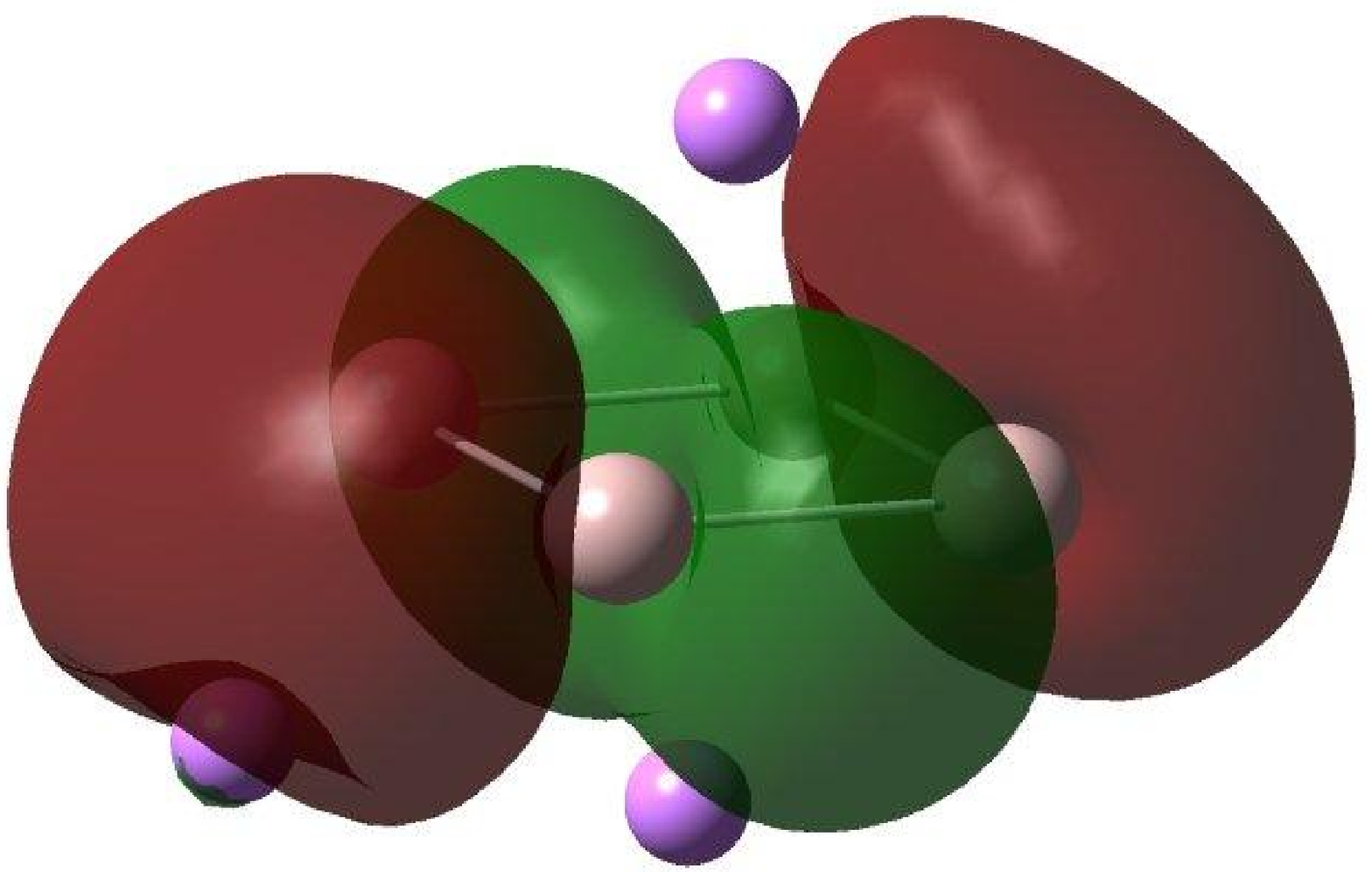}
\end{center}
\end{minipage} \\
(e) & & & & (f) \\
\begin{minipage}[t]{0.25\columnwidth}
\begin{center}
\includegraphics[width=\textwidth]{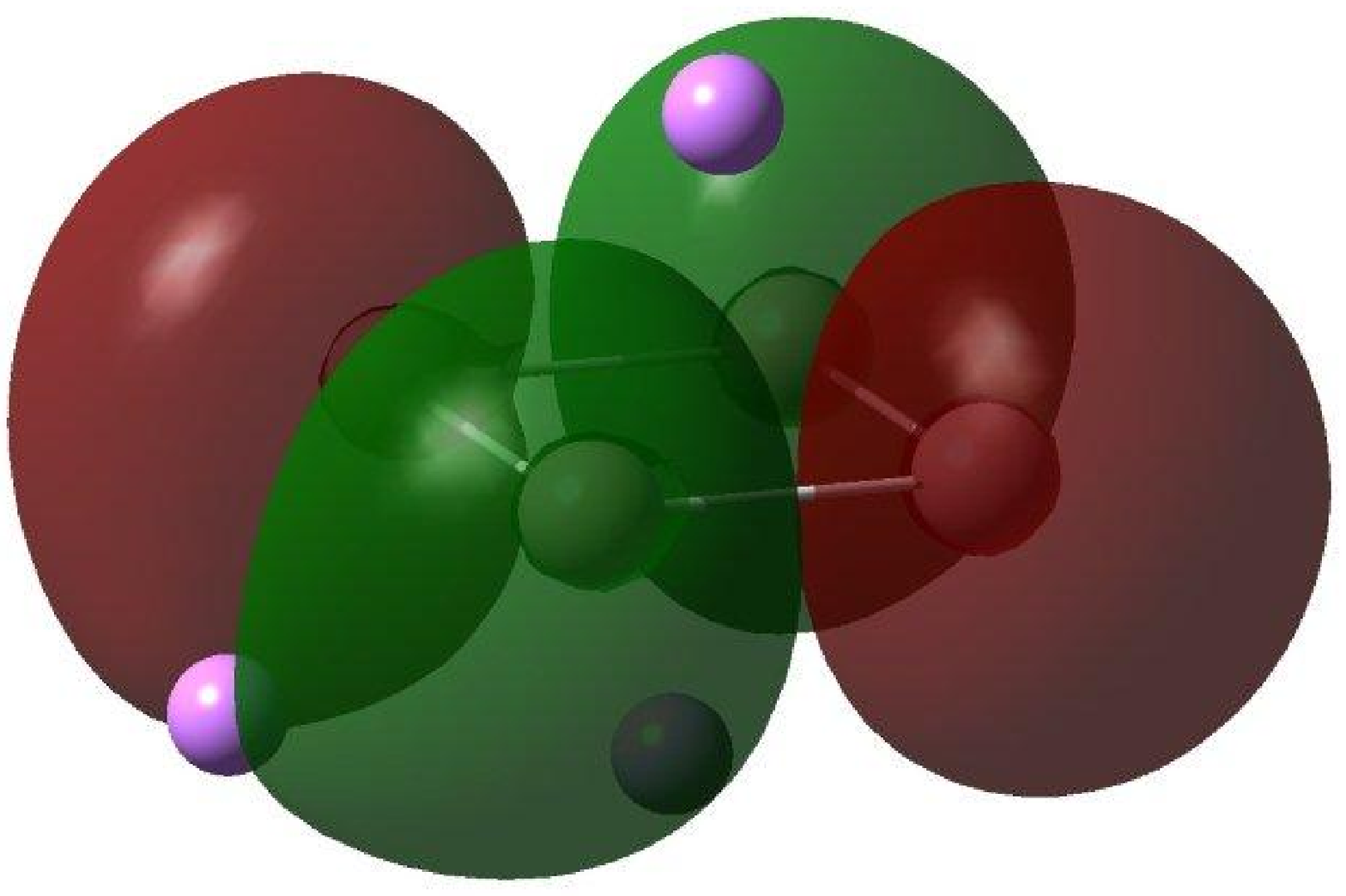}
\end{center}
\end{minipage}
& $-0.10911$ & HOMO $-3$ & $-0.10954$ &
\begin{minipage}[t]{0.25\columnwidth}
\begin{center}
\includegraphics[width=\textwidth]{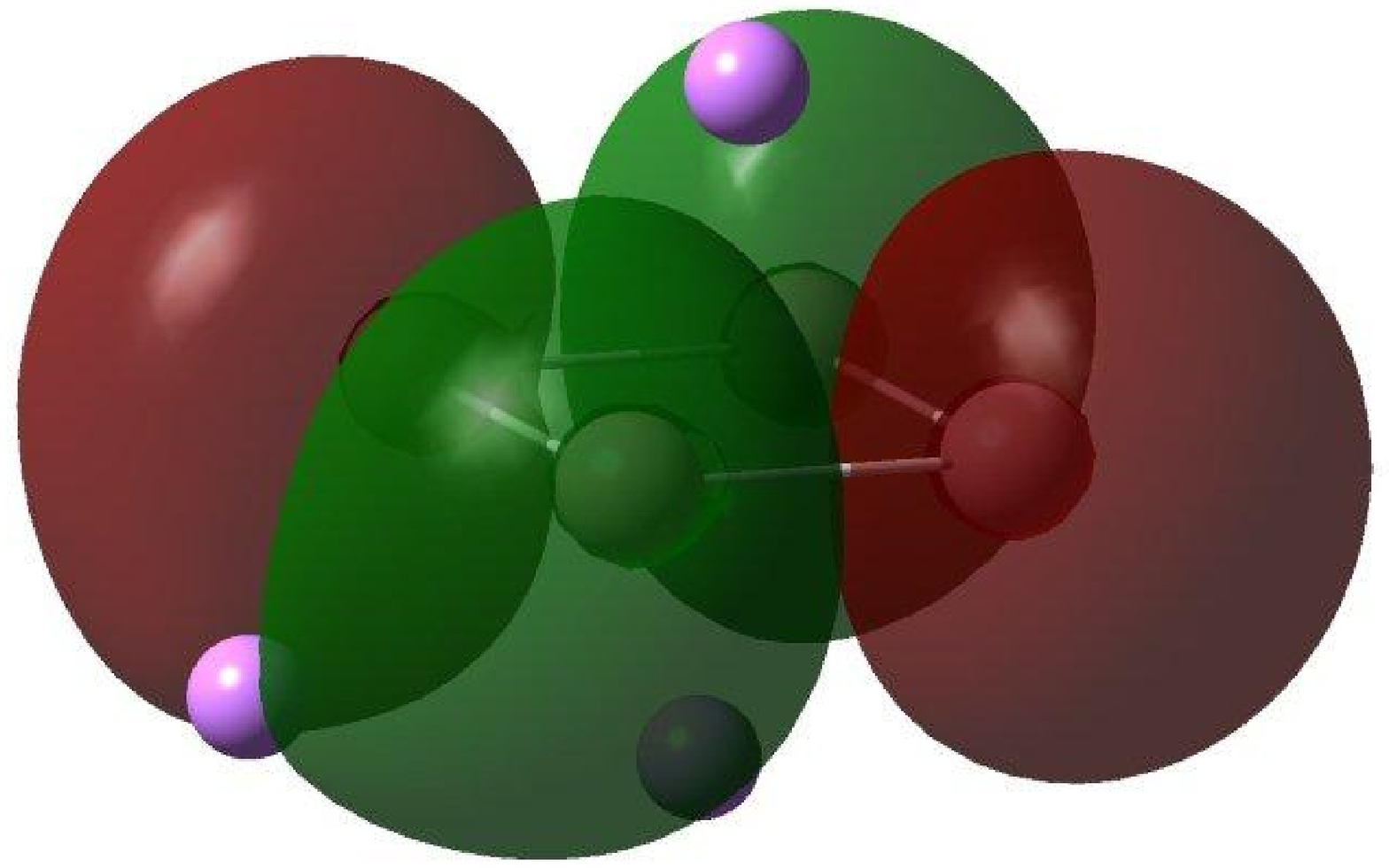}
\end{center}
\end{minipage} \\
(g) & & & & (h) \\
\begin{minipage}[t]{0.25\columnwidth}
\begin{center}
\includegraphics[width=\textwidth]{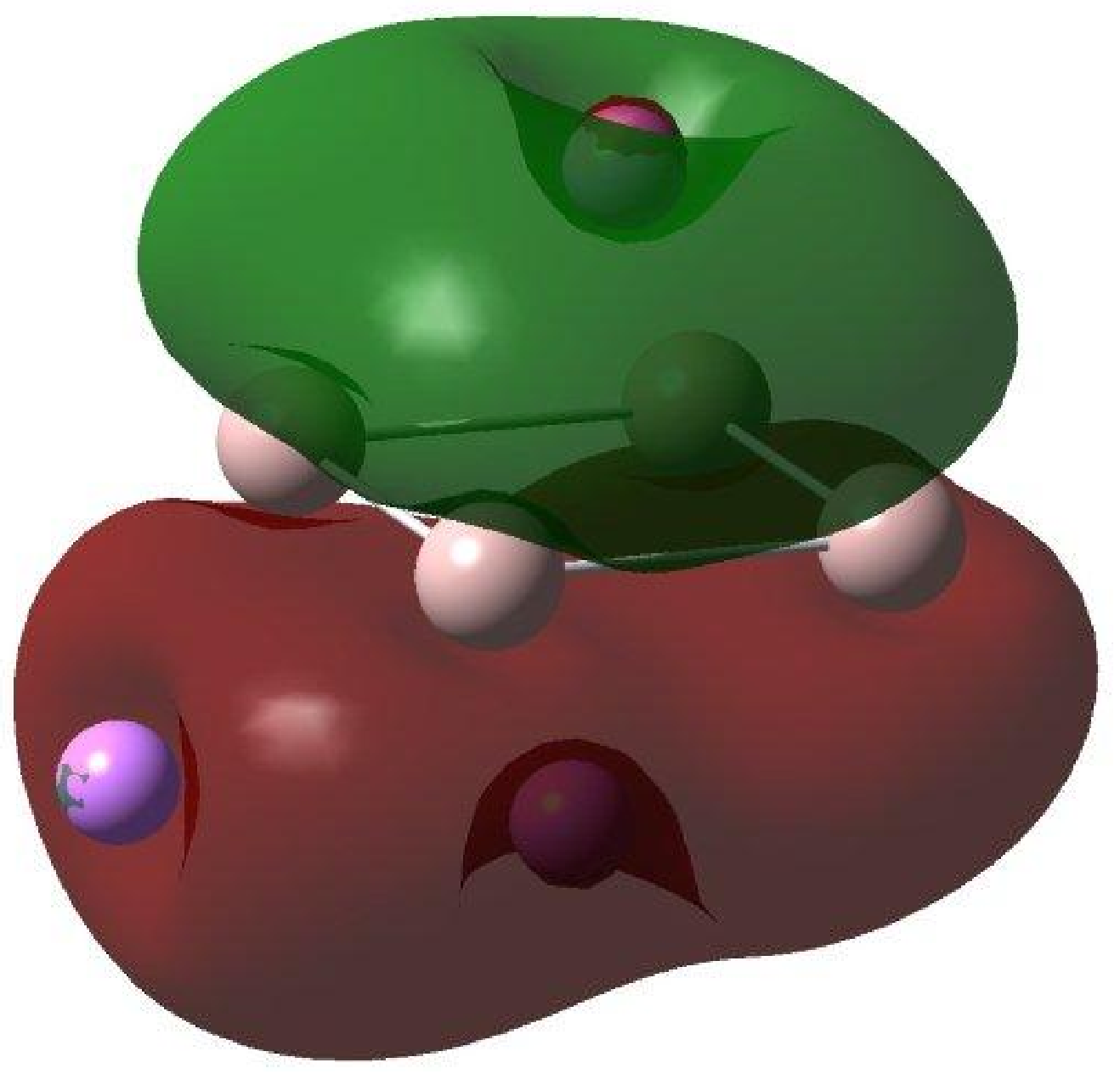}
\end{center}
\end{minipage}
& $-0.12277$ & HOMO $-4$ & $-0.12329$ &
\begin{minipage}[t]{0.25\columnwidth}
\begin{center}
\includegraphics[width=\textwidth]{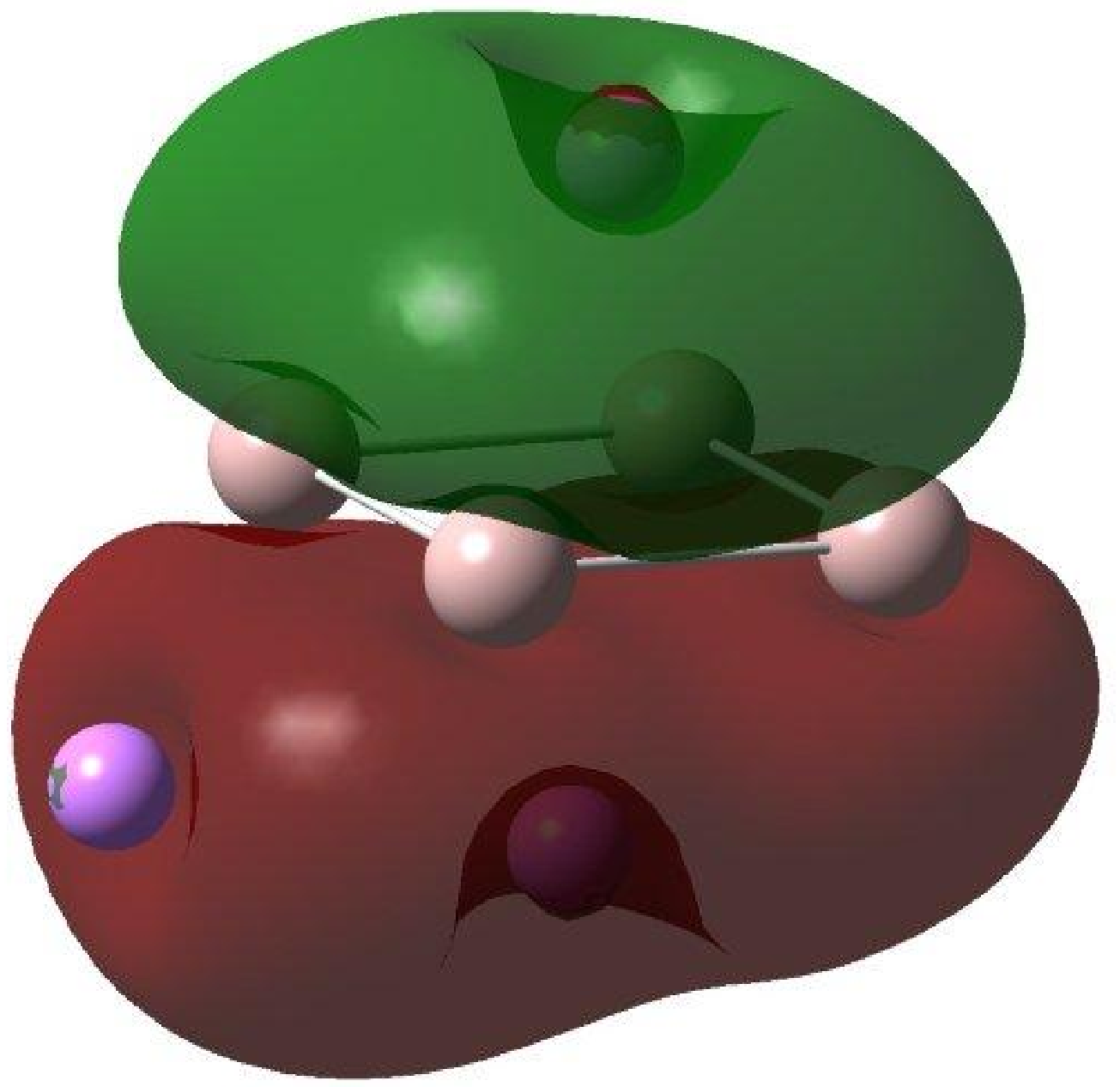}
\end{center}
\end{minipage} \\
(i) & & & & (j) \\
\end{tabular}
\end{center}
\caption{(Color online) HOMO orbitals of the anti-aromatic (left) and aromatic
(right) isomers, and corresponding eigenvalues.}
\label{fig:A1}
\end{figure}

In Fig.~\ref{fig:A1} (left column), we have thought it of interest to record the
shapes of the HOMO orbitals of the anti-aromatic isomer, together with the
corresponding eigenvalues. Fig.~\ref{fig:A1} (right column) shows similar
results for the other low-lying isomer, and it is fair to say that no very clear
fingerprints are in evidence in comparing the two geometries of the negative ion
in Fig.~\ref{fig:A1}

Therefore, we turn immediately to discuss where we have found quite distinct
fingerprints of antiaromaticity, namely in chemical shifts.

\section{Theoretically predicted chemical shifts}
\label{sec:shifts}

This Section finds its motivation in the review of Schleyer \emph{et al.}
\cite{Chen:05}. These authors consider the nucleus-independent chemical shifts
(NICS) as an aromaticity criterion. The development of the original NICS
technique emerged from the study of ring current effects on Li$^+$ chemical
shifts described in the above review.

To match the customary NMR convention, NICS indices correspond to the negative
of the magnetic shielding computed at chosen points in the vicinity of
molecules. NICS is normally computed at ring centres, at points above, and even
at grids in and around the molecule. Significantly negative NICS values in
interior positions of rings or cages indicate the presence of induced diatropic
ring currents (aromaticity), whereas positive indices at each point imply
paratropic ring currents and antiaromaticity. With this brief background, our
results follow.

\begin{table}[t]
\begin{center}
\begin{tabular}{lll}
 & Anti-aromatic & Aromatic \\
 & (rectangle) & (square) \\
$E$ (MP2) (Hartree) & $-990.91594$ & $-990.91340$ \\
 & & \\
 & ppm & ppm \\
Li-1 & 44.0 & $-59.4$ \\
Li-2 & 37.0 & $-82.5$ \\
Li-3 & 102.0 & 102.4 \\
 & & \\
Al-1 & 531.2 & 572.0 \\
Al-2 & 531.2 & 575.9 \\
Al-3 & 719.8 & 734.3 \\
Al-4 & 719.8 & 735.8
\end{tabular}
\end{center}
\caption{Total MP2 energies (Hartree) of ground-state isomers of
(Li$_3$Al$_4$)$^{-1}$, along with their MP2 GIAO Magnetic Shielding in ppm. See
Fig.~\ref{fig:1} for numbering scheme.}
\label{tab:2}
\end{table}

Table~\ref{tab:2} records first the second order M\o{}ller-Plesset (MP2)
energies in Hartrees for the two low-lying isomers of Fig.~\ref{fig:1}. The
antiaromatic case lies lower, though the energy difference is but
0.0025~Hartrees. But the Li shifts are seen from Table~\ref{tab:2} to be
dramatically different from the (almost) rectangular Al$_4$ configuration and
the square geometry. In fact, for the two Li atoms above and below the Al$_4$
plane, namely Li-1 and Li-2, there is a much bigger downfield shift for the
square configuration ($-59.4$ and $-82.5$) than for the rectangular one ($44.0$ and
$37.0$), while for the external one, Li-3, in both cases the shift is around 102,
which is consistent with the value calculated at the same level of the theory
for isolated Li system (Li =101.4;  Li$^{+1}$=93.3; Li$^{-1}$=104.4). 

For the Al atoms the situation is reversed, there is a bigger downfield shift
for the rectangular configuration with a much smaller relative change, compared
to the Li atoms, with respect to the values for the isolated Al system
\footnote{In the case of the neutral Al atom, the MP2 GIAO Magnetic shielding
tensor (ppm) is greatly anisotropic with the following eigenvalues $x =
-19072.8583$, $y = -19072.8583$, $z = 787.2383$. Therefore we consider only the
$z$ component.} (Al = 787; Al$^{+1}$ = 784; Al$^{+2}$ = 776; Al$^{+3}$ = 766).

Notably this is in consilience with the geometric criterion for aromaticity
that the bonding (here around the Al-atom quadrangle) is more delocalized in the
aromatic structure.  The (more stable) anti-aromatic structure has more
localized alternating bonds, as in the case of cyclobutadiene.  

\section{Summary and future directions}

\begin{figure}[t]
\begin{center}
\begin{tabular}{cc}
Anti-aromatic & Aromatic \\
\begin{minipage}[t]{0.45\columnwidth}
\begin{center}
\includegraphics[bb=48 50 1094 866,clip,width=\textwidth]{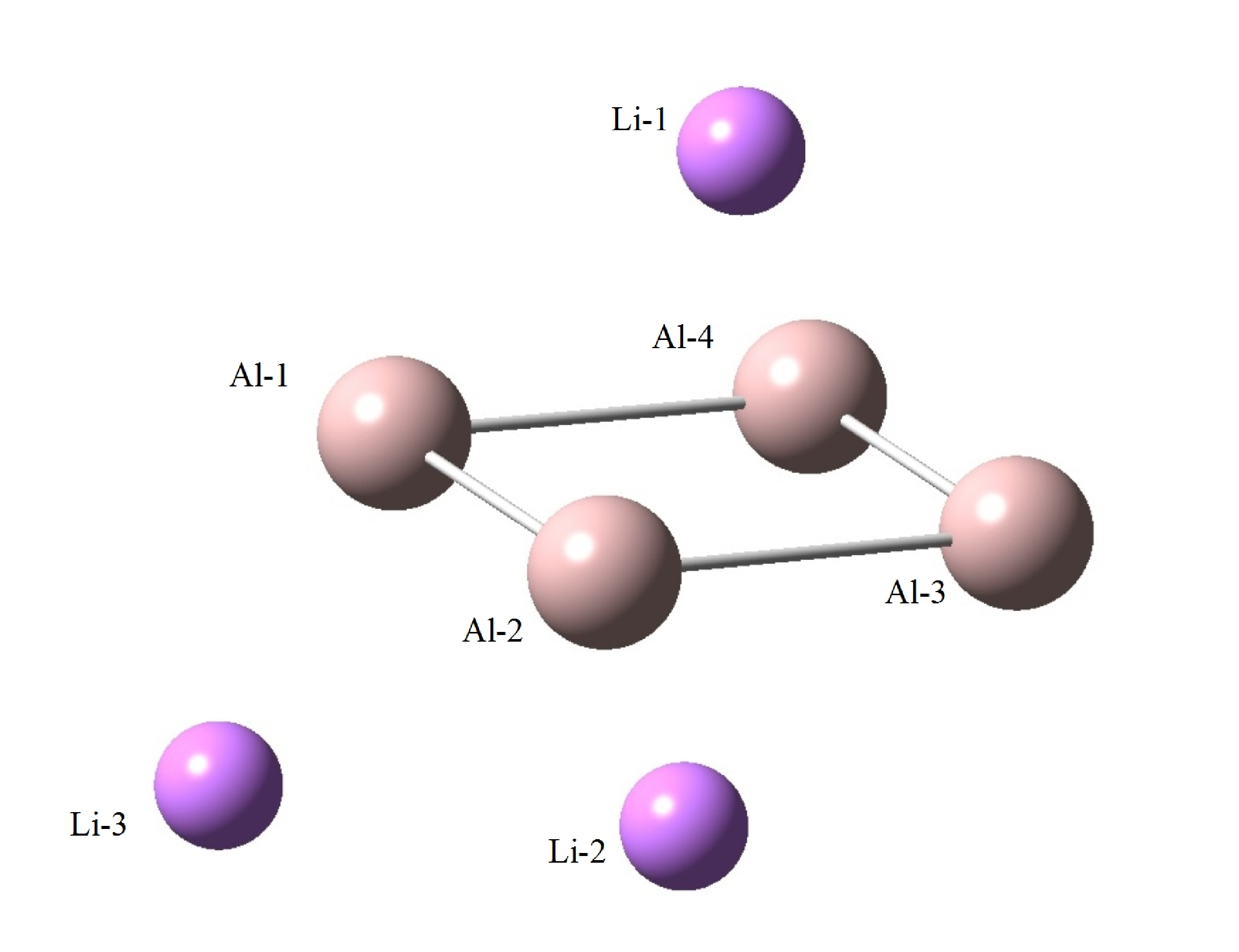}
\end{center}
\end{minipage}
\begin{minipage}[t]{0.45\columnwidth}
\begin{center}
\includegraphics[bb=48 50 1094 866,clip,width=\textwidth]{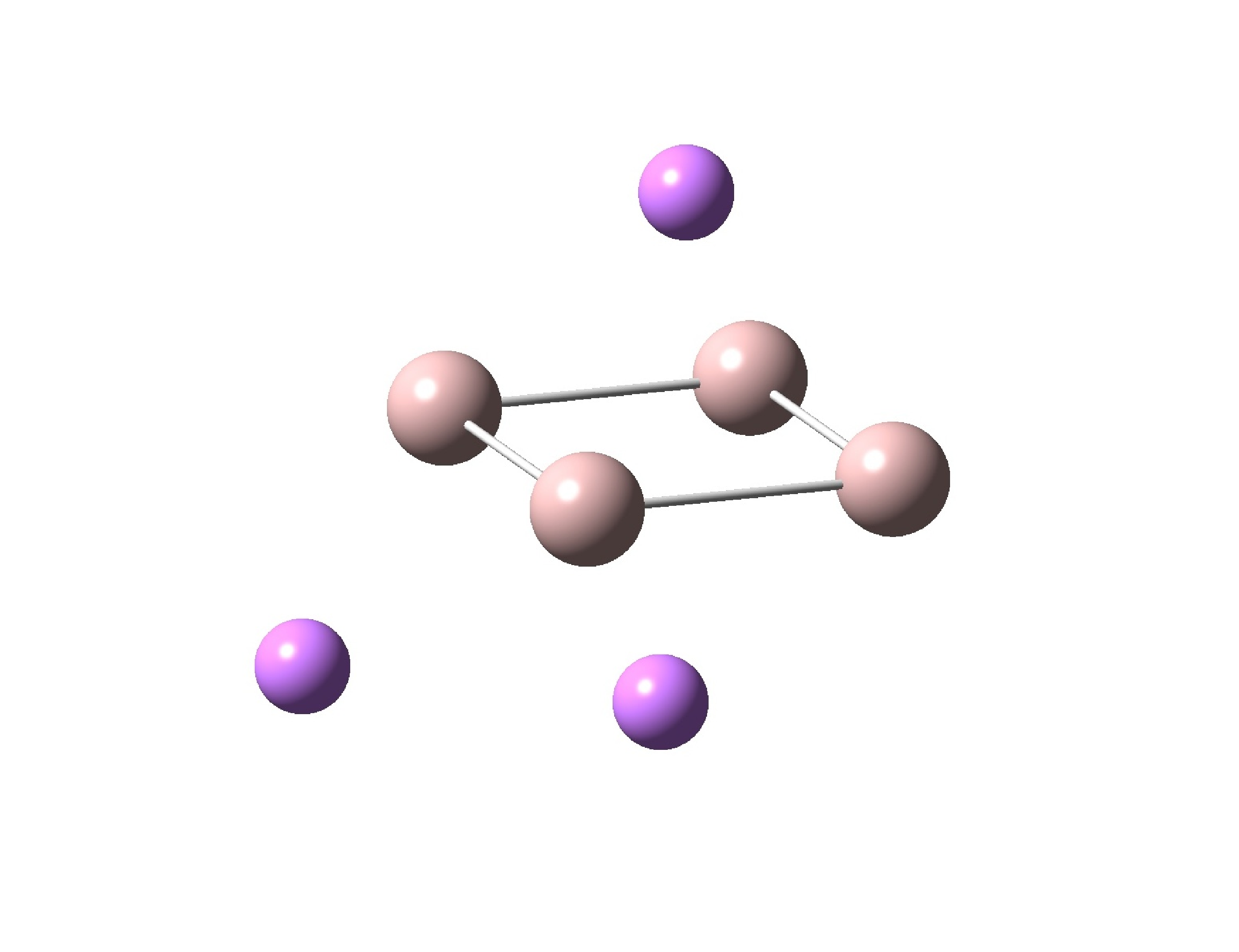}
\end{center}
\end{minipage} \\
(a) rectangle & (b) square
\end{tabular}
\end{center}
\caption{Structure of (Li$_3$Al$_4$)$^{-1}$ and numbering scheme.}
\label{fig:1}
\end{figure}

The important predictions of the present \emph{ab initio} quantum chemical study
are: (i) The geometries of the two low-lying isomers shown in Fig.~\ref{fig:1}.
Here, in Fig.~\ref{fig:1}a the anti-aromatic almost rectangular form of the
Al$_4$ nuclei is shown, while the square Al$_4$ configuration is as predicted in
Fig.~\ref{fig:1}b; (ii) The vibrational frequencies of these two isomers, which
are recorded in Table~\ref{tab:1}; (iii) Most important, the chemical shifts we
predict from the present theoretical investigation are the really sensitive
fingerprints of antiaromaticity, as is clear from Table~\ref{tab:2}.

As to future directions, we want to mention the interest in the Shannon entropy.
This has gained renewed attention by the appearance of the work by Noorizadeh
and Shakerzadeh \cite{Noorizadeh:10}. These authors have proposed a novel
measure of aromaticity based on the local Shannon entropy \cite{March:13} in
information theory. Their index, which measures the probability of electronic
charge distribution between atoms in a given ring, is termed by them the Shannon
aromaticity (SA). These authors observed linear correlations between the SAs
they evaluated by density functional theory (DFT) \cite{Parr:89}, and other
criteria of aromaticity, such as ASE, A and NICS indices, the last of these
being emphasized in our present study on the ion (Li$_3$Al$_4$)$^-$. It is
relevant to record here that Havenith \emph{et al.} \cite{Havenith:04} have
investigated the aromatic character of some Li$_x$Al$_4$ clusters, including
Li$_3$Al$_4$, using ring current patterns.

\section*{Acknowledgement(s)}

NHM wishes to acknowledge that his contribution to the present study was begun
during a visit to the University of Catania. NHM thanks Professors R. Pucci and
G. G. N. Angilella for their generous hospitality during this visit. Also NHM
has continued affiliation with the University of Antwerp (UA) via grant BOF-NOI
(UA), and thanks Professors D. Lamoen and C. Van~Alsenoy for thereby making
possible the continuing affiliation of NHM with UA. Support from the Welch
Foundation of Houston, Texas, is acknowledged through grants BD-0894 (for DJK)
and BD-0046 (for ATB).

\bibliographystyle{gPCH}
\bibliography{a,b,c,d,e,f,g,h,i,j,k,l,m,n,o,p,q,r,s,t,u,v,w,x,y,z,zzproceedings,Angilella,notes}

\end{document}